\documentclass[aps,prb,reprint,showpacs,superscriptaddress,longbibliography]{revtex4-2}
\usepackage{mathtools,amssymb,graphicx,units,bm}

\usepackage[plainpages=false,pdfpagelabels,colorlinks=true,linkcolor=red,urlcolor=blue,citecolor=blue,pdftitle={Title},pdfauthor={},pdfdisplaydoctitle=true,pdfduplex=DuplexFlipLongEdge]{hyperref}
\usepackage{verbatim}
\usepackage[english]{babel}
\usepackage{color}
\newcommand{\tcr}{\textcolor{red}}

\usepackage[normalem]{ulem}

\graphicspath{{../}}

\begin{document}

\title{Bilayer Hubbard model: Analysis based on the fermionic sign problem}
\author{Yingping Mou}
\affiliation{Beijing Computational Science Research Center, Beijing 100193, China}
\author{R. Mondaini}
\email{rmondaini@csrc.ac.cn}
\affiliation{Beijing Computational Science Research Center, Beijing 100193, China}
\author{R.T. Scalettar}
\email{scalettar@physics.ucdavis.edu}
\affiliation{Department of Physics, University of California,
Davis, California 95616, USA}
 
\begin{abstract}
The bilayer Hubbard model describes the antiferromagnet to spin singlet transition and,  potentially, aspects of the physics of unconventional superconductors. Despite these important applications, significant aspects of its `phase diagram' in the interplane hopping $t_\perp$ versus on-site interaction $U$ parameter space, at half filling, are largely in disagreement. Here we provide an analysis making use of the average sign of weights over the course of the importance sampling in quantum Monte Carlo  simulations to resolve several central open questions. Specifically, this metric of the weights clarifies the finite-sized metallic regimes at small $U$. Furthermore, at strong interactions, it points to the existence of a crossover from a correlated to uncorrelated band insulator not yet explored in a variety of existing, unbiased numerical methods. Our paper demonstrates the versatility of using properties of the weights in quantum Monte Carlo simulations to reveal important physical characteristics of the models under study.
\end{abstract}


\maketitle
\section{Introduction} In the wake of the discovery of high-temperature superconductors, and the unlikeliness of their description within a conventional electron-phonon mechanism~\cite{Bednorz1986}, strongly interacting models that could explain the physical mechanisms occurring in cuprates had a surge of investigation. Among these, multilayer geometries such as the bilayer Heisenberg model~\cite{Sandvik1994,Sandvik1996}, are essential to understand the robust (i.e., ~finite temperature) antiferromagnetic ordering observed in undoped materials. The bilayer $t$-$J$~\cite{Lercher1994,Eder1995} and Hubbard models~\cite{Bulut1992,Scalettar1994,dosSantos1994} allowed the study of the interplay of itinerant electrons and (short-ranged) magnetic ordering in the presence of hole doping, and hence spin fluctuation mediated pairing. In the latter, original studies have pointed out the possibility of a nodeless $d$-wave pairing, where the gap has opposite signs in the bonding and anti-bonding Fermi surfaces, and that interplane hybridization weakens in-plane superconducting correlations.

Due to the presence of the sign problem~\cite{Loh1990,Troyer2005,Mondaini2021} in the doped regime~\cite{Rademaker2013}, investigations using quantum Monte Carlo (QMC) simulations had  most success studying the half-filled case~\cite{Bouadim2008,Euverte2013}, which allows the understanding of how global long-range magnetic ordering takes place at sufficiently small interplane hybridizations. More recently, large scale ground-state QMC calculations~\cite{Golor2014} have clarified the absence of metallicity at finite values of the interactions, as initially suggested to occur~\cite{Fuhrmann2006,Kancharla2007,Bouadim2008,Hafermann2009,Ruger2014}, and further corroborated the existence of a magnetic transition in the 3D Heisenberg universality class as the interplane hybridization was increased, similar to that of the bilayer Heisenberg model~\cite{Wang2006}. Dynamical cluster approximation calculations have also examined the possibility of the enhancement of superconductivity as one of the bands approaches the Lifshitz transition, and its implications for heavily electron-doped FeSe-derived superconductors~\cite{karakuzu2021superconductivity}.

In this paper, we revisit the phase diagram of the half-filled bilayer Hubbard model using the finite-temperature determinant quantum Monte Carlo (DQMC) method~\cite{Blankenbecler1981,Hirsch1985}, with a goal of establishing its different phase boundaries in a way that takes the average sign of partial weights in the sampling as a minimal correlator. This metric, recently used to understand criticality in a variety of quantum models~\cite{sign-prob3}, further allows one to unveil a subtle and often unappreciated crossover from a correlated to uncorrelated band-insulating regime at large interplane hybridizations.

\begin{figure*}[t]
    \centering
    \includegraphics[width=0.99\textwidth]{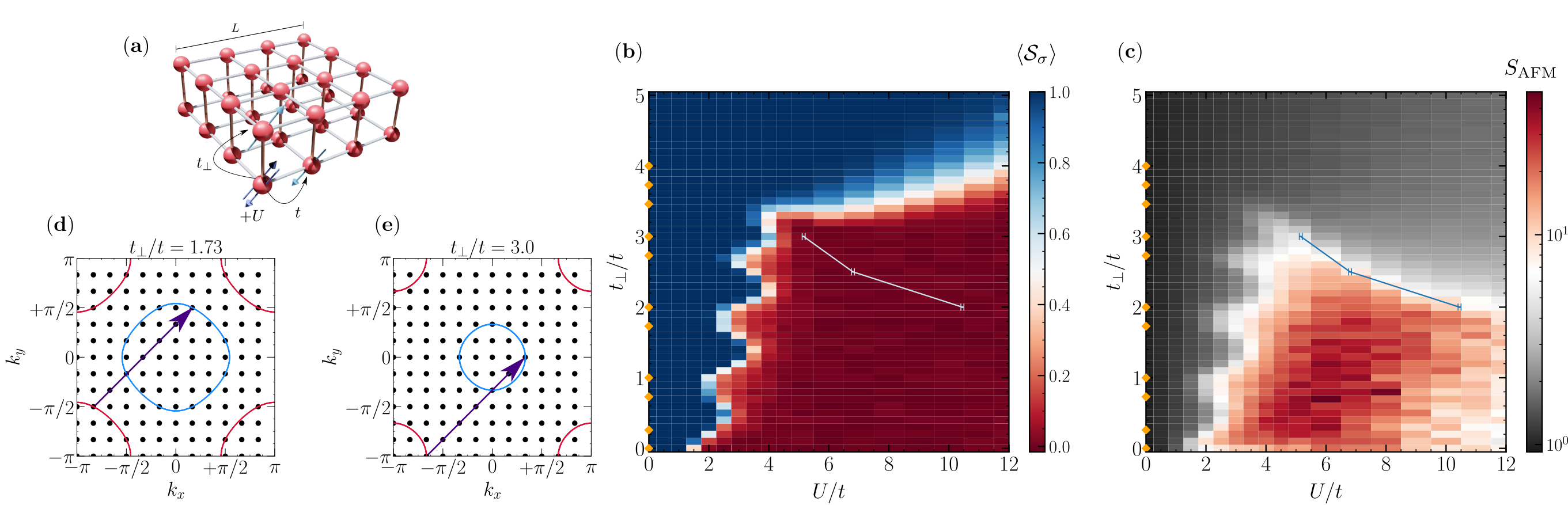}
    \caption{(a) Schematic illustration of the Hamiltonian, with relevant parameters annotated. (b) The contour plot of the average spin resolved sign $\langle{\cal S}_\sigma\rangle$ in the space of parameters $(t_\perp,U)$; (c) the equivalent for the antiferromagnetic structure factor $S_{\rm AFM}$. All data are extracted at temperatures $T/t = 1/20$ in a $12\times12$ bilayer. Markers along the $U/t = 0$ axis in (b) and (c) depict the values of $t_\perp/t$ that result in a nesting condition with a wavevector $(\pi,\pi)$ for this lattice size. Examples of such nestings are shown in (d) and (e) for $t_\perp/t = 1.73$ and 3.0, respectively.Lines in (b) and (c) depict the results of Ref.~\onlinecite{Golor2014} marking the magnetic transition obtained via the scaling of the antiferromagnetic order parameter at $T=0$.}
    \label{fig:fig1}
\end{figure*}

\section{Model} The Hamiltonian of an $L\times L$ bilayer reads 
\begin{eqnarray}
\hat H = 
&-&t \sum_{\langle ij\rangle\ell\,\sigma}  (\hat c^{\dagger}_{i\ell\sigma} \hat c^{\phantom{\dagger}}_{j\ell\sigma} + {\rm H.c.})
-t_\perp \sum_{i \,\sigma}  (\hat c^{\dagger}_{i0\sigma} \hat c^{\phantom{\dagger}}_{i1\sigma} + {\rm H.c.}) \nonumber \\
&+& U \sum_{i\ell} \left(\hat n_{i\ell\uparrow} -\frac{1}{2} \right) 
\left(\hat n_{i\ell\downarrow} - \frac{1}{2}\right)
-\mu\sum_{i,l,\sigma}\hat n_{il\sigma},
\label{eq:ham_spinful}
\end{eqnarray}
where $\hat c^{\dagger}_{i\ell\sigma}$ is the fermionic creation operator on site $i$ of the layer $\ell$ ($\ell = 0, 1$) with spin $\sigma$ ($\sigma=\uparrow,\downarrow$), and $\hat n_{i\ell\sigma}$ is the corresponding number density operator; $t$ and $t_\perp$ quantify the nearest-neighbor intra- and interplane hopping amplitudes, while $U$ is the strength of the local repulsive interactions with chemical potential $\mu$ controlling the electronic density [Fig.~\ref{fig:fig1}(a)].

In the non-interacting limit ($U=0$) at half-filling ($\mu = 0$), the system undergoes a metal-to-band insulator transition as $t_\perp/t > 4$, with a gap opening between the bonding ($-$) and anti-bonding ($+$) bands, $\varepsilon_0^\pm({\bf k})  = -2t[\cos(k_x) +\cos(k_y)] \pm t_\perp$, whose size is $t_\perp - 4t$. In the opposite, strongly interacting, case ($U\gg t,t_\perp$), the Hamiltonian at $\mu = 0$ is equivalent to the bilayer Heisenberg model~\cite{Sandvik1994,Wang2006}, which displays a magnetic transition from an antiferromagnetic ordered bilayer to a quantum disordered phase featuring interplane singlets. Mapping the spin exchange interactions $J$ to the hopping energy scales at this limit, gives a critical hybridization $t_\perp^c/t = 1.588$ separating these two regimes.

For the generic $U\neq0$ case, we solve Eq.\eqref{eq:ham_spinful} by making use of DQMC, in which the introduction of a real-space imaginary-time auxiliary field $\{s_{i\tau}\}$ decouples the interactions, allowing the fermionic integration to be taken exactly. As a result, the partition function is written in terms of the product of weights for each fermionic flavor $\sigma$, ${\cal Z} = \sum_{s} \prod_\sigma w_{\sigma}(\{s_{i\tau}\})$, where the field $\{s_{i\tau}\}$ is being summed. Instead of solving for all configurations $\{s_{i\tau}\}$, importance sampling is performed while observing the convergence of physical observables. The \textit{single}, controllable  approximation used is the imaginary-time discretization $\Delta\tau$ which we take as 0.1 throughout. 

The form of the partition function reveals a peculiarity of the method: the weights being summed are not positive-definite. In fact, in a large class of problems of interest this `sign problem' precludes the accurate computation of physical quantities in the most interesting parts of their phase diagrams~\cite{White89,Loh1990}. At half-filling, however, due to the bipartite structure of the lattice, there is no sign problem~\cite{Hirsch1985}. That is, the product of the signs of the weights in the Monte Carlo sampling is positive regardless of the configuration $\{s_{i\tau}\}$ of the Hubbard-Stratonovich (HS) field. However, even though the \textit{total} sign of the weights is always positive, the sign of individual ones {\it are not}. Indeed, recent results have demonstrated that the average sign of \textit{individual} weights, ${\langle\cal S}_\sigma\rangle \equiv \langle {\rm sgn}(w_\sigma)\rangle$, are directly related to the physics of the Hamiltonian under investigation, following scaling laws similar to those for physical observables in the vicinity of quantum phase transitions~\cite{sign-prob3}. 

\begin{figure*}[t]
    \centering
    \includegraphics[width=1.0\textwidth]{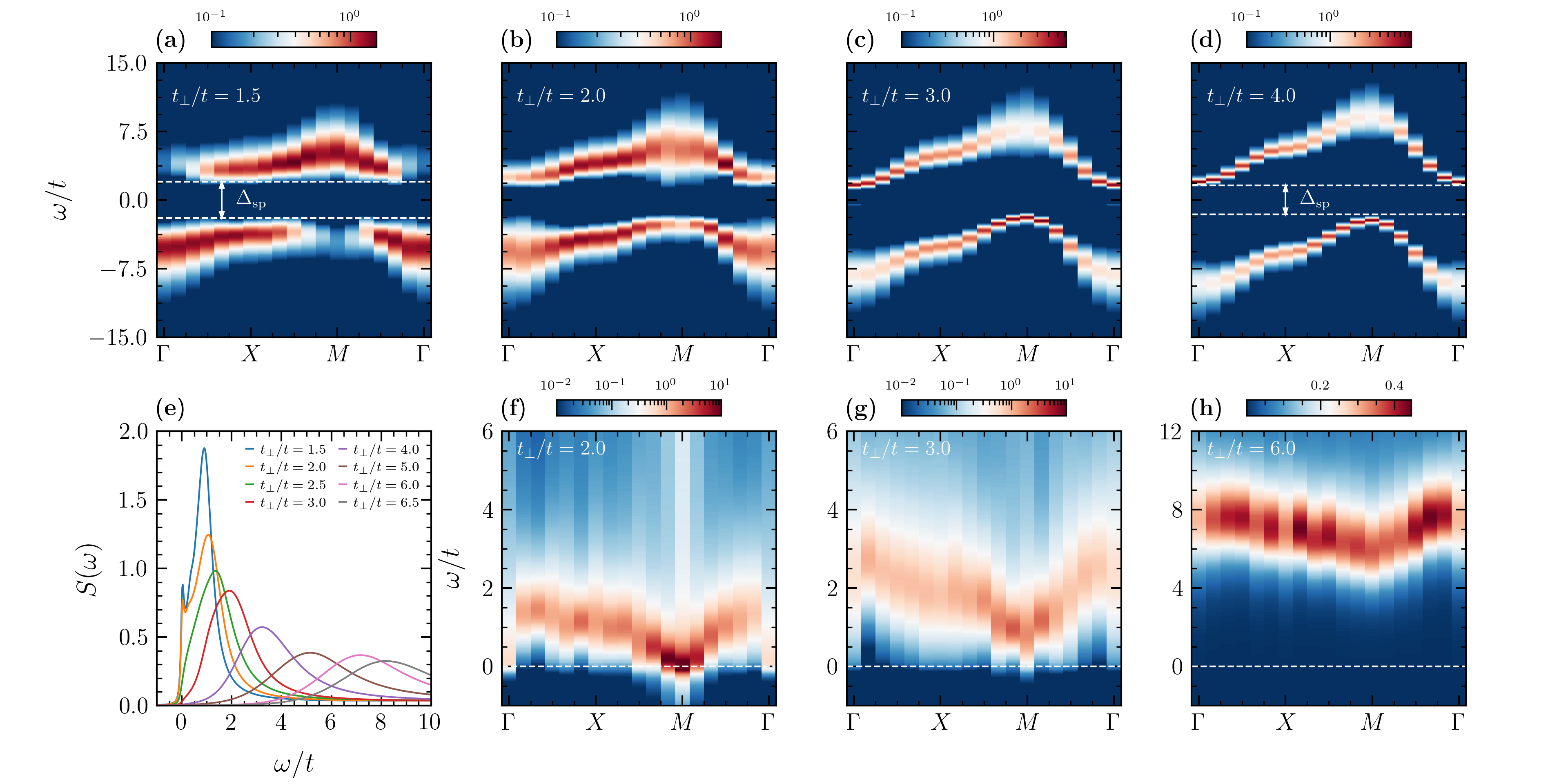}
    \caption{(a)-(d) The single-particle excitation spectrum $A_{\bf k}(\omega)$ across a path connecting high-symmetry points ($\Gamma - X - M - \Gamma$) in the Brillouin zone, for the indicated hybridization amplitudes In (a), $t_\perp/t=1.5$, and (d), $t_\perp/t=4.0$, the single-particle gap $\Delta_{\rm sp}$ is marked, exhibiting, respectively, a direct and indirect behavior, respectively. (f)-(h) Dynamical spin structure factor $S({\bf k}, \omega)$ at different $t_\perp/t$ as marked, with a momentum integrated version shown in (e). Data are obtained at temperatures $T/t = 1/20$ in an $L=12$ bilayer with $U/t=10$.}
    \label{fig:fig2}
\end{figure*}

\section{The sign problem phase diagram}

We start by studying the sign phase diagram of the bilayer Hubbard model in Fig.~\ref{fig:fig1}(b), focusing on the regime $U\lesssim t$. Lobes where $\langle {\cal S}_\sigma \rangle$ is close to 1 appear at small interaction strengths, while in a broad region of parameters $\langle {\cal S}_\sigma \rangle\to0$. Not coincidentally, the boundaries of such lobes in the $U/t\to0$ limit match the loci where a perfect nesting of the Fermi surfaces,  $\varepsilon_0^+({\bf k}+{\bf Q}) = -\varepsilon_0^-({\bf k})$, occur in a finite lattice [see Figs.~\ref{fig:fig1}(d) and \ref{fig:fig1}(e) for two examples] when spanning the interplane hybrization $t_\perp/t$~\cite{Golor2014}.
These lobes are finite-size effects which vanish in the thermodynamic limit, where the spacing between nesting conditions similarly vanishes other system sizes are shown in Appendix~\ref{app:other_sizes}. Moreover, the ${\bf Q} = (\pi,\pi)$ nesting suggests that the system is unstable toward antiferromagnetic order, and ensuing Mott insulating behavior is the most probable scenario, as further pointed out by mean-field calculations~\cite{Golor2014}.
The immediate conclusion is that the regions at small $U/t$ with $\langle {\cal S}_\sigma \rangle \simeq 1$ identify the metallic regime observed in \textit{finite} lattices over a variety of numerical methods~\cite{Fuhrmann2006,Kancharla2007,Bouadim2008,Hafermann2009,Ruger2014}.

Going beyond small interactions, studies that use $\langle {\cal S}_\sigma \rangle$ to track the phase transitions~\cite{sign-prob3} have shown that the region with a vanishing spin-resolved sign can be identified, in related models and when approaching the thermodynamic limit, with either the magnetically ordered regime or the one yielding a Mott insulator. This is the case for the SU(2) Hubbard model in the honeycomb lattice, for example, where both phases are known to concomitantly occur~\cite{Paiva05, Meng10, Sorella12}. Here, in the SU(2) bilayer Hubbard model, however, by overlaying in Fig.~\ref{fig:fig1}(b) the accurate numerical results from Ref.~[\onlinecite{Golor2014}] for the critical interaction strength $U_c(t_\perp)$ that leads to the onset of magnetism in the $T\to0$ limit, we demonstrate that $\langle {\cal S}_\sigma \rangle \to 0$ is not tracking the magnetic phase transition. Our results resolve the transition (or crossover) it  probes, and, in doing so, provide insight into the transition between uncorrelated and correlated band insulating regimes.

This becomes more evident if contrasting Figs.~\ref{fig:fig1}(b) and \ref{fig:fig1}(c). By plotting the antiferromagnetic structure factor $S_{\rm AFM} \equiv (1/2L^2)\sum_{i,j} (-1)^{i+j}\langle(\hat n_{i\uparrow} - \hat n_{i\downarrow})(\hat n_{j\uparrow} - \hat n_{j\downarrow})\rangle$, in a relatively large lattice ($L=12$) [Fig.~\ref{fig:fig1}(c)], the critical points $U_c(t_\perp)$ systematically border the regime where $S_{\rm AFM}$ is large. The average spin-resolved sign, on the other hand, departs from zero at larger interplane hybridizations [Fig.~\ref{fig:fig1}(b)]. In particular, that the magnetic transition is not accompanied by a Mott transition in the bilayer Hubbard model was initially shown within cluster dynamical mean-field (DMFT) calculations~\cite{Kancharla2007}, which demonstrated the existence of a \textit{paramagnetic} Mott insulator preceding the onset of a band insulating state at larger $t_\perp$~\footnote{Exact diagonalization results in Appendix~\ref{app:ED} provide further evidence that the magnetic and the Mott insulating transitions are not aligned at large interaction strengths in the bilayer Hubbard model.}. 

\section{The Mott insulator-Band insulator crossover}

Differentiating between Mott and band insulators at sufficiently large $t_\perp$ is challenging. This difficulty has been illustrated not only in theoretical studies of model Hamiltonians, but also in the experimental characterization of certain transition-metal dichalcogenides~\cite{Wang2020}, which exhibit a competition of on-site Coulomb repulsion and interlayer hopping in its layered structure. As both phases naturally manifest a finite gap $\Delta_{\rm sp}$ for single-particle excitations, either separating the upper and lower Hubbard bands in the Mott phase or the bonding and anti-bonding bands in the band insulating regime, a useful distinguishing characteristic is provided by the trend of $\Delta_{\rm sp}$ with growing interplane hybridization~\cite{Kancharla2007} as well as the direct or indirect nature (in momentum) of $\Delta_{\rm sp}$.

To implement this approach, we start by reporting the single-particle spectral function $A_{\bf k}(\omega)$ in Fig.~\ref{fig:fig2}, calculated via the stochastic analytic continuation of imaginary-time dependent QMC data~\cite{Sandvik1998},
\begin{equation}
    {\bm G}({\bf k},\tau) = \int \frac{d\omega}{\pi} \frac{e^{-\omega \tau}}{1+e^{-\beta\omega}} A_{\bf k}(\omega),
\end{equation}
where ${\bm G}({\bf k},\tau)$ is the space Fourier transform of the imaginary time ($\tau$) displaced Green's function ${\bm G}({\bf k}, \tau) = \langle \Psi_{{\bf r + R}, \sigma}(\tau)\Psi^{\dag}_{{\bf r},\sigma}(0)\rangle$, where the creation operator is given as $\Psi^{\dag}_{{\bf r},\sigma} = ({\hat c}^{\dag}_{{\bf r}0,\sigma}, {\hat c}^{\dag}_{{\bf r}1,\sigma})$ with the creation operator of the electron at unit cell ${\bf r}$ and sublattice $0, 1$ with spin $\sigma$.

\begin{figure}[t]
    \centering
    \includegraphics[width=0.99\columnwidth]{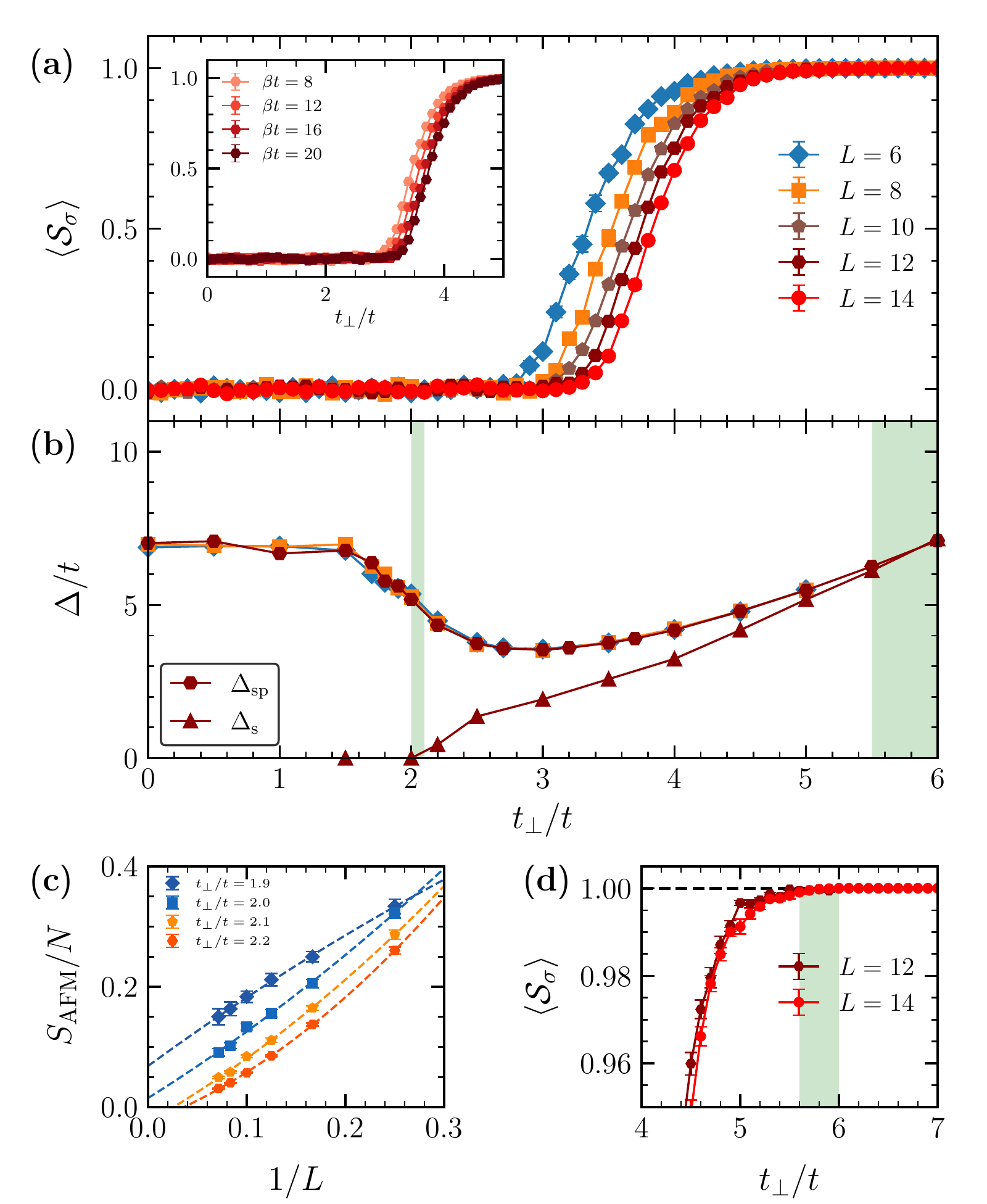}
    \caption{(a) A cut along $U/t = 10$ of the spin-resolved average sign $\langle {\cal S}_\sigma\rangle$ with growing interplane hopping $t_\perp$ and various system sizes. The inset highlights the significant drift in the upturn location $t_\perp/t$ of $\langle {\cal S}_\sigma\rangle$ as one decreases the temperature $T$ (or increases the inverse temperature $\beta=1/T$) for an $L=12$ bilayer. (b) The corresponding single-particle gap $\Delta_{\rm sp}$ extracted by the minimum gap of single-particle excitations across the whole Brillouin zone, at comparable system sizes; also shown the gap to spin excitations $\Delta_{\rm s}$, finite once the AFM order is absent. This is also demonstrated by the vanishing of the normalized (by the system size) AFM structure factor, shown in (c), in the thermodynamic limit at $t_\perp = 2.05(5)t$ see first vertical shaded region in (b). The second shaded region at $t_\perp = 5.75(25)t$ describes the onset of $\Delta_{\rm sp} \simeq \Delta_s$, also seen in the zoom-in of the saturation of $\langle {\cal S}_\sigma\rangle$ in (d). Data are extracted at temperatures $T/t = 1/20$. }
    \label{fig:fig3}
\end{figure}

By focusing on large interactions, $U/t=10$, we notice that at small $t_\perp$ [$t_\perp \lesssim 2t$, see Fig.~\ref{fig:fig2}(a)], the single-particle excitation bands display a direct gap at the $X=(\pi,0)$ point of the Brillouin zone as in the single-layer Hubbard model, whereas for larger interlayer hoppings ($t_\perp \gtrsim 2t$), the gap becomes indirect, connecting the $\Gamma$-$M$ [$(0,0)$-$(\pi,\pi)$] points instead, Figs.~\ref{fig:fig2}(c) and \ref{fig:fig2}(d). Moreover, for values of $t_\perp \approx 3t$, the indirect gap reaches a minimum, whereupon increasing hybridization leads to a larger $\Delta_{\rm sp}$. This is summarized in Fig.~\ref{fig:fig3}(b) for different system sizes additional details in the $A_{\bf k}(\omega)$ results are described in Appendix~\ref{app:spectral}, and an independent scheme (with similar results) that bypasses the analytic continuation is given in Appendix~\ref{app:direct_extract_gap}.

This analysis allows us to divide the dependence of $\Delta_{\rm sp}$ on the hybridization at fixed interaction strength in three classifications: (i) a gap that is largely robust to the increase of $t_\perp$ up until $t_\perp/t \simeq 1.5-2$, (ii) a decrease of the gap with $t_\perp$ as the hybridization approaches $t_\perp/t \simeq 3$, and (iii) a growth of $\Delta_{\rm sp}$ for values $t_\perp/t \gtrsim 3$. The observed trend of the gap $d\Delta_{\rm sp}/dt_\perp<0$ ($d\Delta_{\rm sp}/dt_\perp>0$) was used before~\cite{Kancharla2007} in cluster-DMFT results to discern the Mott (band-insulating) phases. Consequently, the three regimes we describe can be interpreted, respectively, as the antiferromagnetic Mott insulator, paramagnetic Mott insulator~\footnote{We are currently unaware of other possible orders that may manifest in the described paramagnetic Mott insulator. A rung bond-ordered wave is not a good candidate -- see Appendix~\ref{app:bow}.}, and band insulator. Noticeably, quantitatively similar conclusions about the gap change and the Mott-to-band insulator transition were obtained in smaller clusters using dynamical cluster approximation~\cite{Lee2014}, albeit at more modest interactions ($U/t=6$).

\section{Characterizing band insulators} 

The classification of the onset of a band insulator via the evolution of $\Delta_{\rm sp}$ with $t_\perp$ is however incomplete. It has been argued that a true (or uncorrelated) band insulator is characterized by identical gaps for particle and spin excitations~\cite{Sentef2009}. To verify this, we compute the dynamical spin structure factor, $S(\textbf{k},\omega)$ via the inversion of the integral equation,
\begin{equation}
   \langle \hat{S}({\bf k},\tau)\hat{S}({\bf -k},0)\rangle = \int \frac{d\omega}{\pi} \frac{e^{-\tau\omega}}{1-e^{-\beta\omega}} \chi''({\bf k},\omega),
\end{equation}
where $\chi^{\prime\prime}({\bf k},\omega)$ is the dynamical spin susceptibility, which allows the extraction of $S({\bf k}, \omega) = \chi^{\prime\prime}({\bf k},\omega)/(1-e^{-\beta\omega})$.
Figures~\ref{fig:fig2}(f)-\ref{fig:fig2}(h) reports the dynamical spin structure factor for a representative path in the Brillouin zone. A finite spin gap $\Delta_s$ appears at values $t_\perp/t \gtrsim 2$ at the $M = (\pi,\pi)$ point, in agreement with the disappearance of antiferromagnetic long-range order, as also seen in the $1/L$ extrapolation of the normalized equal-time antiferromagnetic spin structure factor, $S_{\rm AFM} = \frac{1}{2L^2}\sum_{i,j} {\rm exp}\{{\rm i}(\pi,\pi)\cdot({\bf r}_i - {\bf r}_j)\}  \langle (\hat n_{i,\uparrow}-\hat n_{i,\downarrow}) (\hat n_{j,\uparrow}-\hat n_{j,\downarrow})\rangle$ [Fig.~\ref{fig:fig3}(c)]. Larger interplane hybridizations lead to a much reduced momentum dependence in $S({\bf k}, \omega)$; a momentum-integrated version is shown in Fig.~\ref{fig:fig2}(e), and a compilation of the spin gap extracted from $S(\omega)$ is presented in Fig.~\ref{fig:fig3}(b). Both gaps, $\Delta_{\rm sp}$ and $\Delta_s$, acquire similar values at $t_\perp/t \approx 5.8$, thus marking the onset of the \textit{uncorrelated} band-insulating phase.

We are now in position to analyze how $\langle {\cal S}_\sigma\rangle$ captures different crossovers, as reported in Fig.~\ref{fig:fig3}(a). Despite not insignificant system-size dependence, these data indicate that the upturn of the average spin-resolved sign at sufficiently low temperatures occurs close to the Mott insulator-band insulator crossover at around $t_\perp \approx 3t$ [as originally seen in Fig.~\ref{fig:fig1}(b)]. In turn, $\langle {\cal S}_\sigma\rangle$ saturates at one precisely when the uncorrelated band-insulating regime takes place. Figure \ref{fig:fig3}(d) shows a detailed zoom-in of this $\langle {\cal S}_\sigma\rangle\to1$ approach. Such saturation is also seen in quantifiers of the typically sampled fields in DQMC, e.g.,~the average Hamming distance~\cite{Tiancheng2021} --see Appendix~\ref{app:Hamming}. Although a scaling analysis based on the average sign of the weights $\langle {\cal S}_\sigma\rangle$~\cite{sign-prob3} is elusive, likely due to the absence of an intrinsic order parameter characterizing either phase, other models where a band-insulating phase takes place, as the square lattice ionic Hubbard model~\cite{Garg06,Paris07,Bouadim07,Craco08,Garg14}, also exhibit a convergence of $\langle {\cal S}_\sigma\rangle$ towards one, as we similarly observe here~\cite{Mondaini2021,sign-prob3}. Further characterization of this crossover is shown in Appendix~\ref{app:local_corr} which describes several local spin correlators.

\noindent\section{Summary and outlook} 
By using a combination of quasiparticle excitation gaps with the average sign of the weight of one fermionic flavor $\langle {\cal S}_\sigma\rangle$ in the importance sampling of DQMC simulations, we identified the different phases of the bilayer Hubbard model at half filling. Beyond the clarification of the system size-influenced metallic regimes at small interactions, one of our main results using this tracker is the onset of an uncorrelated band-insulating regime at large interplane hybridizations. While this transition is likely a crossover, i.e., without an associated order parameter, the saturation of $\langle {\cal S}_\sigma\rangle$ at one coincides with the regime where spin and single-particle excitations are comparable. As a result, in both uncorrelated phases at finite lattices, metallic and band insulating, $\langle {\cal S}_\sigma\rangle\to 1$. That $\langle {\cal S} \rangle$ might signal transitions in model Hamiltonians is consistent with its role as a necessary ingredient to compute \textit{any} physical observable~\cite{Hirsch1985}, thus inherently revealing details of the physics at play. Moreover, for the specific case of the crossover between band insulating phases, it has the added advantage that it does not require the quantification of, often expensive, time-displaced correlation functions, neither analytic continuation of the data.

A testament of the relevance of these results is provided by experiments involving ultracold atoms trapped in optical lattices, which have recently succeeded in emulating the bilayer Hubbard model~\cite{Koepsell2020,Gall2021}, opening a further, and highly controllable, realization for precision investigations of its different phases.

Going beyond half filling, our exploration of the sign problem in this model may allow one to tackle the regime which is mostly relevant to the doped cuprates, where bilayer features have been observed~\cite{Feng01} and numerically interpreted~\cite{Liechtenstein96}, culminating in the investigation of the potential onset of a finite-temperature superconducting transition governed by a Kosterlitz-Thouless form. Such analysis has been recently carried out for the case of a single layer~\cite{sign-prob3}. Its extension to a bilayer model might reveal the impact of the interplane hybridization on the finite critical temperature $T_c$, as well as on the dominant pairing channel before this transition takes place~\cite{Lanata2009,Maier2011,Matsumoto2020,Bohrdt2022}. 

\section{Acknowledgements} Y.M.~was supported by Grant No. 12047507 funded by the Natural Science Foundation of China (NSFC). R.M.~acknowledges discussions with H.~Guo, and support from the NSFC Grants No. U1930402, No. 12050410263, No. 12111530010, No. 11974039, and No. 12222401. R.T.S.~was supported by Grant DE‐SC0014671 funded by the U.S.~Department of Energy, Office of Science. Illuminating discussions with M.~Randeria are gratefully appreciated. Computations were performed with the QUEST toolbox~\cite{QUEST} on the Tianhe-2JK at the Beijing Computational Science Research Center.

\appendix

\begin{figure}[b]
    \centering
    \includegraphics[width=1.0\columnwidth]{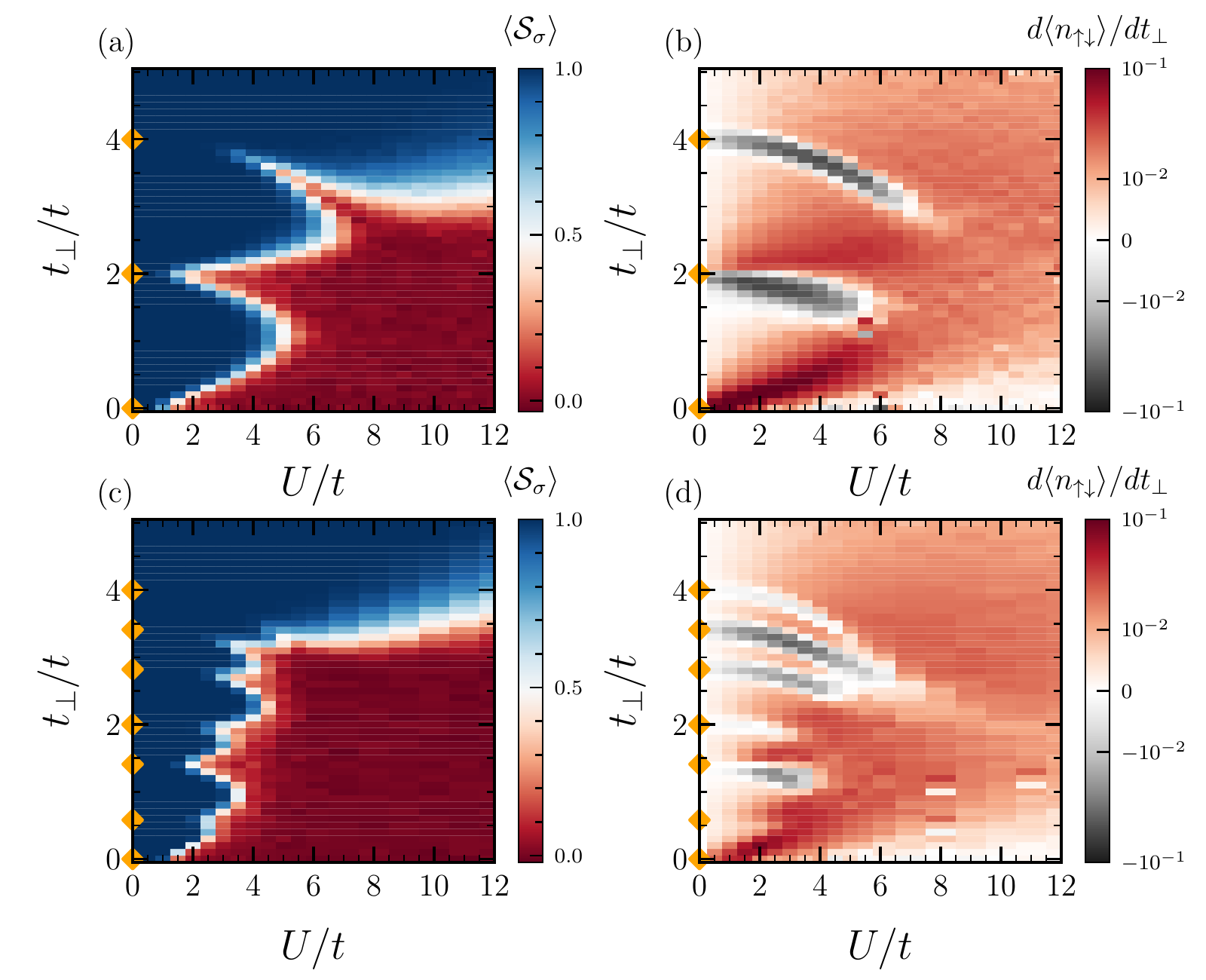}
    \caption{ (a) [(b)] The average spin-resolved sign $\langle {\cal S}_\sigma\rangle$ [derivative of the double occupancy with respect to $t_\perp$, $d\langle \hat n_{\uparrow\downarrow}\rangle/dt_\perp$] on the $T$ versus $U/t$ map for a $4\times4$ bilayer. (c) and (d) show the same for an $L=8$ lattice. As in 
    Fig. \textcolor{red}{1}, markers along the $U/t = 0$ axis denote nesting conditions for the corresponding lattice size, and data are extracted at temperature $T/t = 1/20$.}
    \label{fig:figS_size}
\end{figure}

\section{Local quantities and different system sizes} \label{app:other_sizes} 

From the discussion presented in the main text, the location of the system size-influenced metallic regions is already evidenced by the regimes where the $\langle {\cal S}_\sigma\rangle$ converges to one. To make this connection clearer, and observe how they extend as the lattice size is reduced, we show in Figs.~\ref{fig:figS_size}(a) and \ref{fig:figS_size}(c), the `sign phase diagram' for $L=4$ and 8, respectively. The lobe boundaries in the noninteracting limit are again given by the values of $t_\perp/t$ where a nesting condition occur, even more clearly as the number of such matches is proportional to the system size. A systematic shrinking of these regions takes place with growing $L$.

While the weight of the configurations in the sampling might initially be thought to be an artifact of the quantum simulation algorithm, it directly captures the behavior of physical observables, including the double occupancy, $\langle \hat n_{\uparrow\downarrow}\rangle \equiv (1/2L^2)\sum_i\langle \hat n_{i\uparrow} \hat n_{i\downarrow} \rangle$. In an insulator-to-metal (metal-to-insulator) transition, the double occupancy increases (decreases) with changes in the parameter driving the evolution, for example, the interplane hybridization.  Figures~\ref{fig:figS_size}(b) and \ref{fig:figS_size}(d) show that the derivative of the double occupancy with respect to $t_\perp$, $d\langle \hat n_{\uparrow\downarrow}\rangle/dt_\perp$, reflects this prediction, and identifies the metallic region boundaries in a finite lattice, in direct agreement with the average sign analysis.

\begin{figure}
    \centering
    \includegraphics[width=1.0\columnwidth]{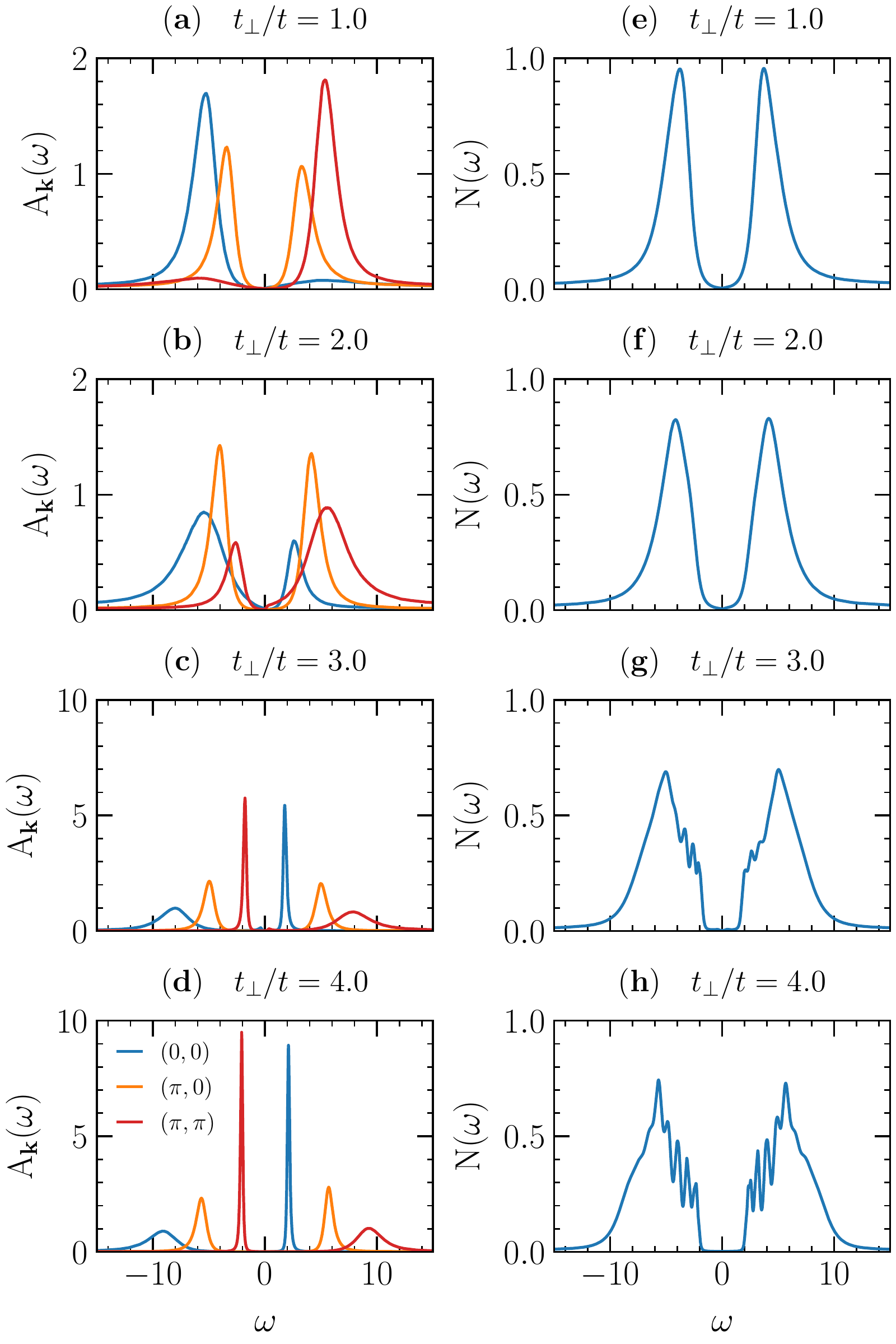}
    \vspace{-0.6cm}
    \caption{The spectral function $A_{\bf k}(\omega)$ (a)-(d) and the single-particle density of states
    $N(\omega)$ (e)-(h) for various interplane hoppings as marked. For values $t_\perp/t \gtrsim 2$ the gap turns indirect, connecting the two bands at the conduction and valence bands at $(\pi,\pi)$ and $(0,0)$, respectively. The density of states at large $t_\perp/t$'s displays a collection of features connected to the incoherent superposition of interplane singlet-states. Data is obtained for $U/t = 10$, $T/t = 1/20$ in an $L=12$ bilayer.}
    \label{fig:figS_dos}
\end{figure}

\section{\textbf{QMC results: Spectral function and DOS}}\label{app:spectral}

Following the discussion in the main text that establishes that the magnetic transition is accompanied by a transition from a direct-to-indirect gap of the single-particle excitations, we display in Figs.~\ref{fig:figS_dos}(a)-\ref{fig:figS_dos}(d) the momentum-resolved $A_{\bf k}(\omega)$ for a range of $t_\perp$ values. To contrast these results, we further show the full density of states, 
$N(\omega) = \sum_{\bf k}A_{\bf k}(\omega)$ in Figs.~\ref{fig:figS_dos} (e)-\ref{fig:figS_dos}(h).

For the spectral function, the change of the gap type, from direct to indirect, is seen to happen at values $t_\perp/t \gtrsim 2$, with a direct $[(\pi,0) - (\pi,0)]$ gap giving way to $[(\pi,\pi) - (0,0)]$, closely following the magnetic transition at $U/t=10$~\cite{Golor2014}. Deep in the nonmagnetic phase the $(\pi,\pi)$ contribution to the excitations departs from the Fermi energy, and thus becomes less relevant at low energies. In turn, the density of states, which is always gapped when increasing the interplane hoppings, displays a variety of incoherent peaks that are a characteristic of the superposition of singlet-states formed across the bilayer in the regime $t_\perp/t\gg1$. In the main text, Fig.
\ref{fig:fig3}(b), the gaps were extracted via the energy difference given by the location of the maximum values of the $\delta$-like functions closest to the Fermi energy at $\omega=0$.

\begin{figure}
    \centering
    \includegraphics[width=1.0\columnwidth]{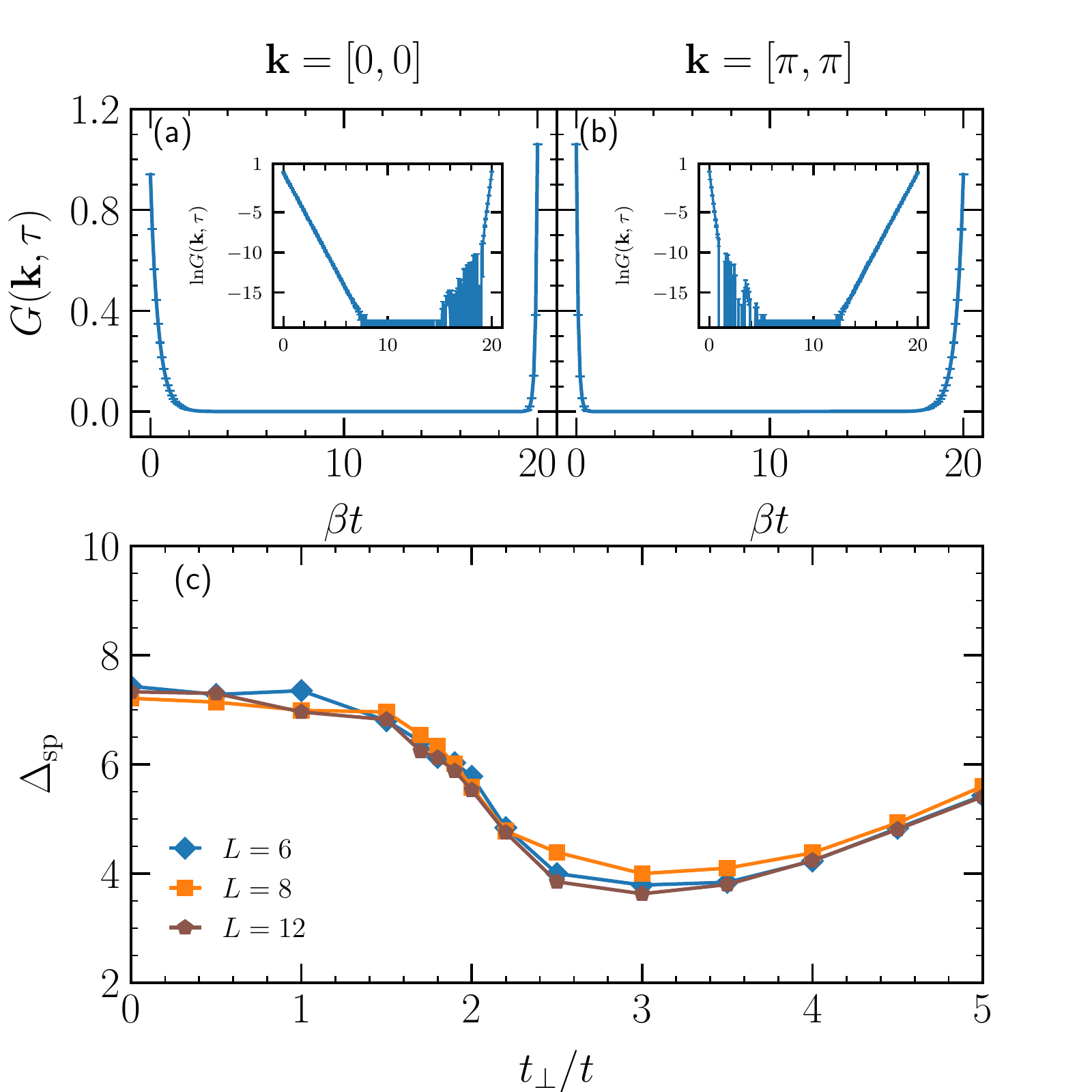}
    \caption{(a),(b) The imaginary-time ($\tau$) dependent Green's function $G(\bf k, \tau)$ at $U/t=10$, $t_{\perp}/t=4.5$ for $L=12$. Inset: $G({\bf k}, \tau)$ on semi-logarithmic scales. (c) The single-particle gap $\Delta_{\rm sp}$ extracted by the minimum gap of fitting ${\rm ln} G({\bf k}, \tau) \sim \pm \omega_\mp({\bf k}) \tau$ across the whole Brillouin zone (see text).}
    \label{fig:fig_gap}
\end{figure}

\section{Directly extracting the single-particle gap}\label{app:direct_extract_gap}

In the main text, we make use of stochastic analytic continuation to invert the integral equation [Eq.~(\tcr{2})] that allows one to retrieve the spectral function $A_{\bf k}(\omega)$ (and consequently the single-particle gap) via the examination of the smallest gap across the allowed momentum points, $\Delta_{\rm sp}({\bf k})$. Here, we employ a direct approach that complements this procedure. The single-particle gap can be similarly obtained by noticing that $G({\bf k}, \tau) \propto e^{\pm \omega_\mp({\bf k}) \tau}$, for imaginary-time $\tau$ sufficiently away from its limits at $\tau=0$ and $\tau = \beta$. In particular, for the case of $\tau <\beta/2$ ($\tau >\beta/2$) one obtains the gap $\omega_+$ ($\omega_-$) above (below) the Fermi energy at $\mu = 0$. As a result, the single-particle gap is extracted as $\Delta_{\rm sp} = \min_{\bf k}[\omega_+({\bf k})] + \min_{\bf k}[\omega_-({\bf k})]$. Figures~\ref{fig:fig_gap}(a) and \ref{fig:fig_gap}(b) exemplify the $\tau$ dependence of the imaginary-time displaced Green's functions at two momentum points, ${\bf k} = (0,0)$ and $(\pi,\pi)$, respectively. A least-squares fitting of the exponential form close to its ends reveals the gap sizes, and also an indirect gap between these two momenta forms at values of $t_\perp/t = 4.5$, as reported in the main text. Compiling these results, across a range of interplane hybridization values at $U/t=10$ [Fig.~\ref{fig:fig_gap}(c)], leads to a $\Delta_{\rm sp}$ largely consistent with the one originally presented in Fig.~\tcr{4}(b), with a larger size dependence, however. The minimum at $t_\perp/t \simeq 3$ still locates the transition from a paramagnetic Mott insulator to the band-insulating regime.

\section{Analysis of a possible bond-order wave}\label{app:bow}

The energy gaps displayed in Fig.~\ref{fig:fig3}(b) in the main text gave us indications for the existence of multiple phases with growing $t_\perp$. In particular, a possible candidate to explain the intervening regime in between the AFM Mott insulator and the band-insulator at large interplane hybridization is a rung bond-ordered wave (BOW). To account for this possibility, we calculate the BOW order parameter,
\begin{equation}
    \langle \hat B_{i} \rangle \equiv \sum_{\sigma}(\hat c^{\dagger}_{i0\sigma}\hat c^{\phantom{\dagger}}_{i1\sigma}+\hat c^{\dagger}_{i1\sigma}\hat c^{\phantom{\dagger}}_{i0\sigma}),
\end{equation}
where, in the notation presented in the main text, the second sub-index refers to the fermionic operator layer.
Verification of the existence of long-range order is accomplished via monitoring the BOW structure factor, $S_{\rm BOW}=(1/L^2)\sum_{ij} e^{ -{\rm i} {\bf k}\cdot ({\bf r}_i - {\bf r}_j) }\langle \hat B_{i} \hat B_{j} \rangle$. 

\begin{figure}
    \centering
    \includegraphics[width=0.65\columnwidth]{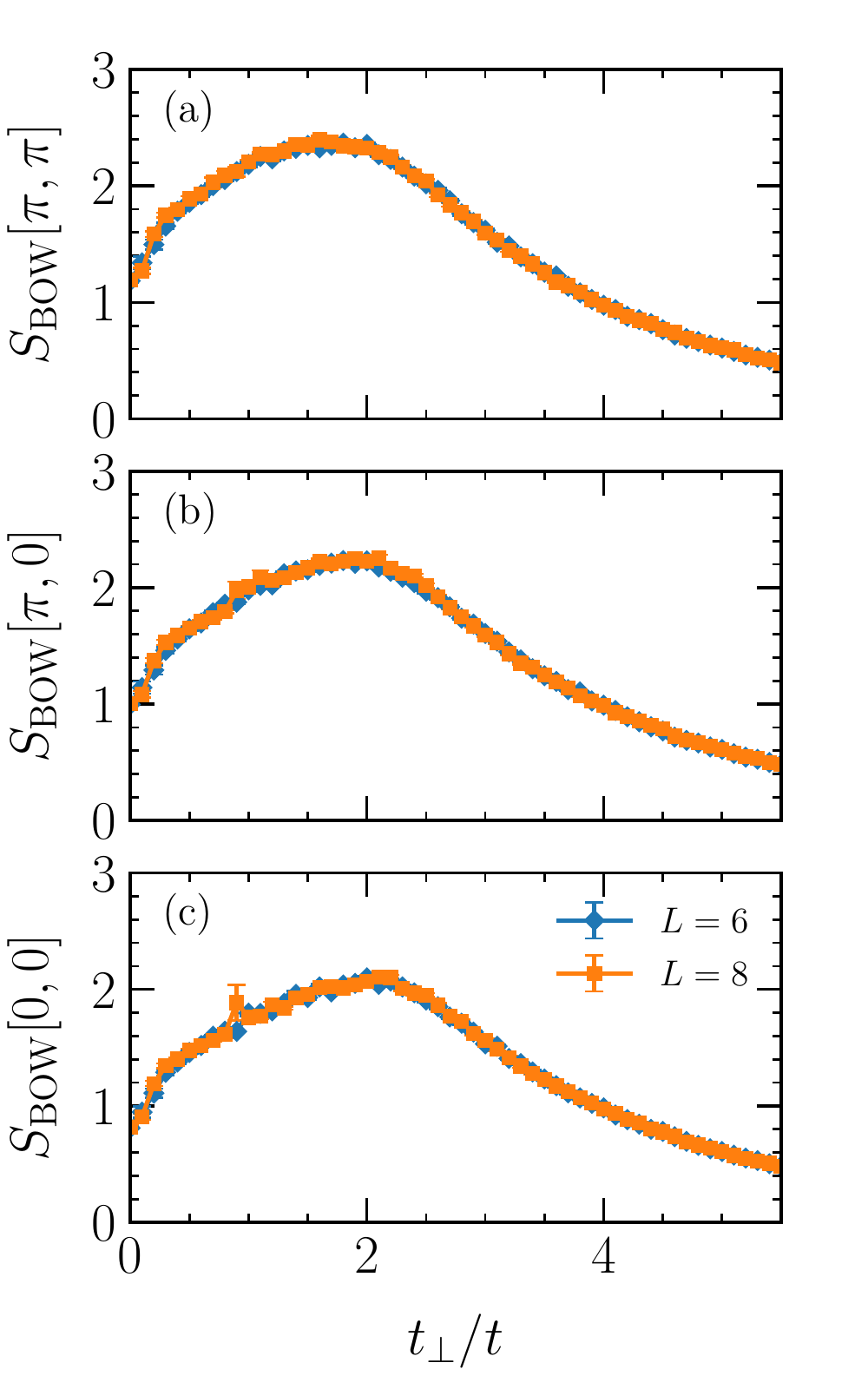}
    \caption{BOW structure factor $S_{\rm BOW}$ as a function of the interlayer hopping $t_{\perp}/t$ for two system sizes at $U/t=10$ and $T/t = 1/20$. Different symmetry channels are displayed, (a) ${\bf k = [\pi, \pi]}$, (b) $[\pi, 0]$, and (c) $[0,0]$. The absence of any significative size dependence indicates there is no long-range bond order.
    }
    \label{fig:fig_bow}
\end{figure}

In Fig.~\ref{fig:fig_bow}, the structure factors are shown as a function of the interlayer hopping for two system sizes, as obtained for a system with $U/t=10$. In all channels we investigate, ${\bf k} = [\pi,\pi], [0,\pi]$ or $[0,0]$, no apparent size dependence of the BOW structure factor can be observed for a large range of interplane hybridizations, suggesting the absence of such ordering in the thermodynamic limit.

\section{Local correlators}\label{app:local_corr}

\begin{figure}
    \centering
    \includegraphics[width=1.0\columnwidth]{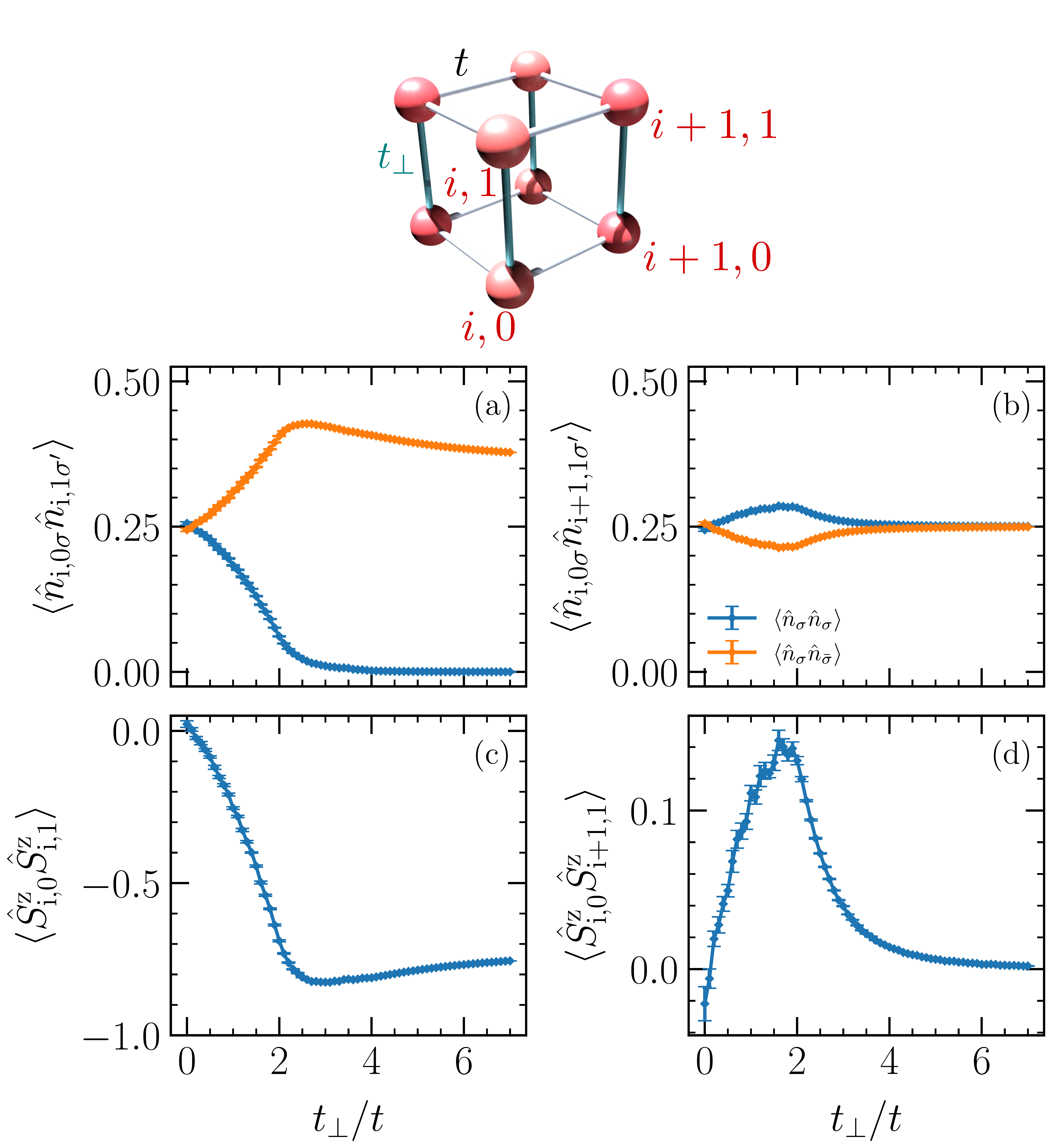}
    \caption{ Top: Cartoon depicting the different sites $(i,l)$ where the two-point correlations are computed as displayed in the lower panels. Bottom: The interlayer (a),(b) density-density correlations for same ($\sigma = \sigma^\prime$) and different spin components ($\sigma \neq \sigma^\prime= - \sigma$) within the same unit-cell (a) and nearest-neighbor unit cells (b). (c) and (d) display the equivalent for the spin-spin correlations. All data  are extracted at $U/t=10$ for $L=14$ and $T/t=1/20$. Spin summation is implicitly assumed for the density correlations.}
    \label{fig:fig_obs}
\end{figure}

While BOW rung correlations are short-ranged, other local correlators aid in drawing a picture of the low-energy physics as the interplane hopping is increased. The density-density correlations within a unit cell [Fig.~\ref{fig:fig_obs}(a)] start at the value 0.25 when the planes are uncoupled ($t_\perp = 0$) and evolve  such that equal-spin correlations are quickly suppressed whereas opposite spin ones are enhanced until $t_\perp/t \simeq 2.1$. Accompanied by the large negative interplane, intra-unit cell spin-spin correlations [Fig.~\ref{fig:fig_obs}(c)], these point to a robust interplane singlet formation in this regime, as a local signature of the overall antiferromagnetic state. Past this threshold in the hybridization $t_\perp$, $\langle \hat S_{i,0}^z \hat S_{i,1}^z\rangle$ decreases in magnitude, signaling the long-range order is absent and interplane singlets in different unit cells become more and more independent. This can be seen explicitly in Figs.~\ref{fig:fig_obs}(b) and \ref{fig:fig_obs}(d), which report the interplane two-point correlations (density and spin, respectively) for nearest-neighbor unit cells: they asymptotically approach their uncorrelated values, pointing to the decreasing interdependency of the different singlets within a unit cell. 
This approach is continuous (at the low, but finite temperatures $T/t = 1/20$) and does not indicate a sharp transition to the collective product state of singlet states. Rather, they are suggestive of a crossover, which is behind the explanation of the smooth transition from the correlated to uncorrelated band-insulating phases.

\section{Hamming Distance} \label{app:Hamming} 

Recent investigations in various fermionic models have shown that the onset of criticality can be directly tracked by quantities related to metrics of the auxiliary HS field~\cite{Tiancheng2021,Natanael2021}, bypassing the necessity of extraction of physical observables. This field, local to each orbital in the spinful Hubbard model in its $(d+1)$-dimensions formulation~\cite{Blankenbecler1981}, is the quantity being sampled via a Metropolis algorithm over the course of the Monte Carlo sampling~\cite{Hirsch1985}. A useful metric (although others exist~\cite{Tiago2021a, Tiago2021b}) is given by the average distance traveled with respect to a given point in the phase space of configurations $\{s_{i,\tau}\}$, and can be defined by the $L_1$-distance (or Hamming distance) of configurations,
\begin{equation}
    {\cal HD} = \frac{1}{2\cal V} \sum_{i, \tau} |s_{i,\tau} - s_{i,\tau}^{\rm ref.}|,
    \label{eq:HD_SU2_def}
\end{equation}
with respect to a reference configuration $\{s_{i,\tau}^{\rm ref.}\}$. Here ${\cal V}$ is the length of the field, $L_\tau \cdot 2L^2$ in our case, with $L_\tau$ being the number of imaginary-time slices that discretize the inverse temperature: $\beta/L_\tau = \Delta\tau$. Assuming that $\{s_{i,\tau}^{\rm ref.}\}$ is a typical configuration (namely, one obtained after a significant number of warmup sweeps in the field), storing ${\cal HD}$ after each sweep on the sampling allows one to quantify the average distance the Markov chain probes in the $(d+1)$-dimensional phase space. This was shown to be intrinsically related to the physics of the models being investigated~\cite{Tiancheng2021}. For example, unordered phases were demonstrated to be associated with completely uncorrelated HS configurations, rendering an average Hamming distance, $\overline {\cal HD}$, equal to 0.5. Departure from this value signals correlation within the sampled space, and thus physically ordered phases.

\begin{figure}
    \centering
    \includegraphics[width=1.0\columnwidth]{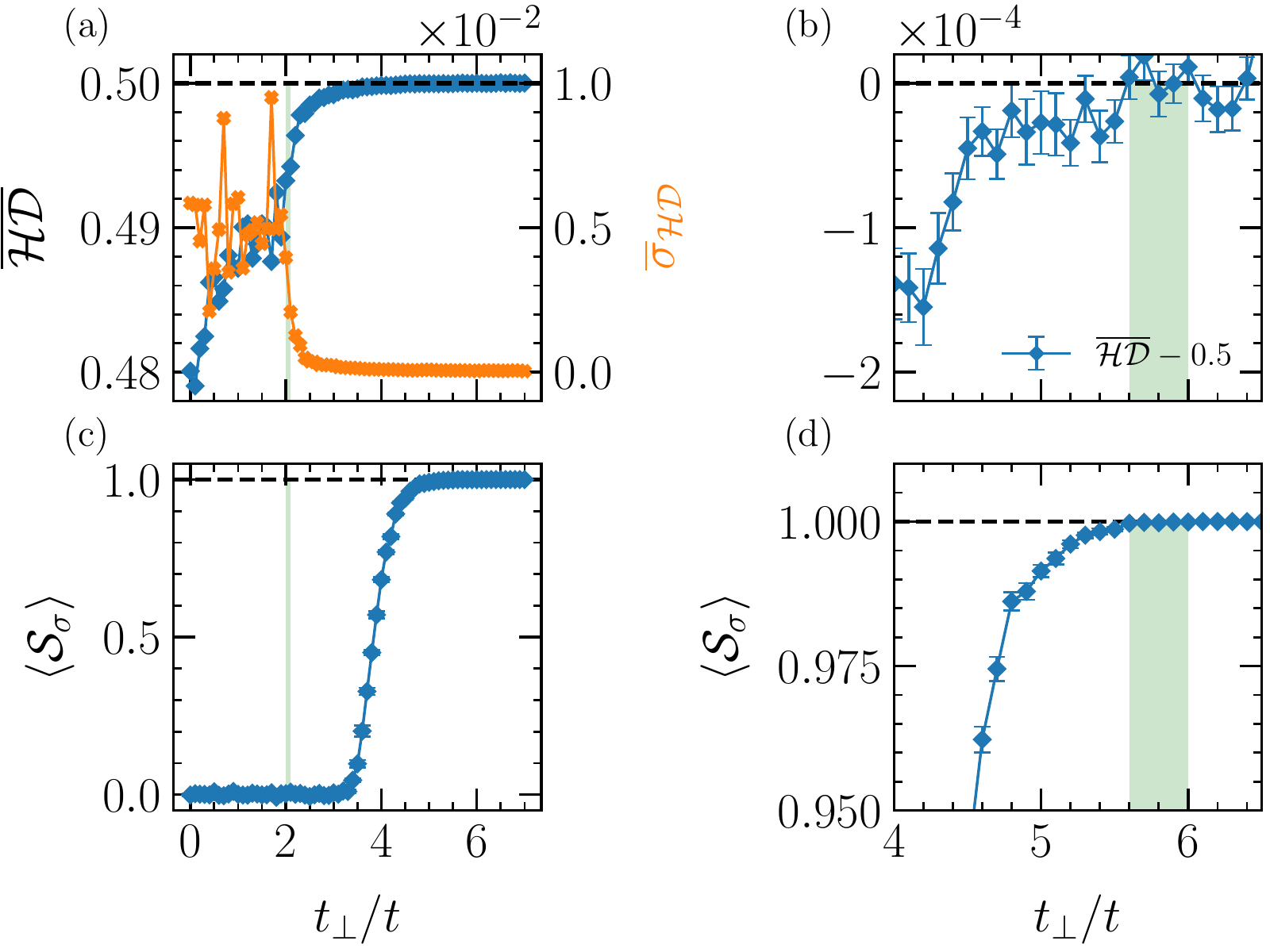}
    \caption{(a) Average Hamming distance and averaged standard deviation of Hamming distances versus ~interplane hybridization with $U/t=10$ and $L=14$. (c) The corresponding spin-resolved sign $\langle {\cal S}_\sigma\rangle$. Vertical shaded regions in (a) and (d) display the QCP obtained by AFM structure factor $t^{c}_{\perp}/t \approx 2.1$, whereas the shaded regions in the zoomed-in panels (b) and (d) mark the confidence region in $t_\perp$ that indicates the correlated-to-uncorrelated band insulating crossover (see main text).}
    \label{fig:fig_hd}
\end{figure}

Applying this idea to the bilayer Hubbard model, we report in Fig.~\ref{fig:fig_hd} the average Hamming distance $\overline {\cal HD}$ as a function of the interplane hybridization at large interactions $U/t = 10$. A significant departure from the uncorrelated sampling occurs at $t_\perp \approx 5.8t$, which marks the regime where the \textit{uncorrelated} band-insulator gives way to a correlated one [see Fig.~\ref{fig:fig3}(b) in the main text]. Another metric of the sampling is the typical width $\sigma_{\cal HD}$ of the Gaussian-distributed Hamming distances within a given realization. Averaging among independently seeded Markov chains, $\overline{\sigma}_{\cal HD}$, results in a marked location for the change of the distributions, which coincides with the onset of the ordered phase at $t_\perp/t \approx 2.1$. 
That is, Hamming distances obtained at each sampling process are much more diverse within the ordered phase in comparison to the ones within the physically unordered regime. These results lend extra insight about the location of the different phases, and have been checked to exhibit qualitatively small finite-size effects.

\section{ED results}\label{app:ED}

\begin{figure}
    \centering
    \includegraphics[width=1.0\columnwidth]{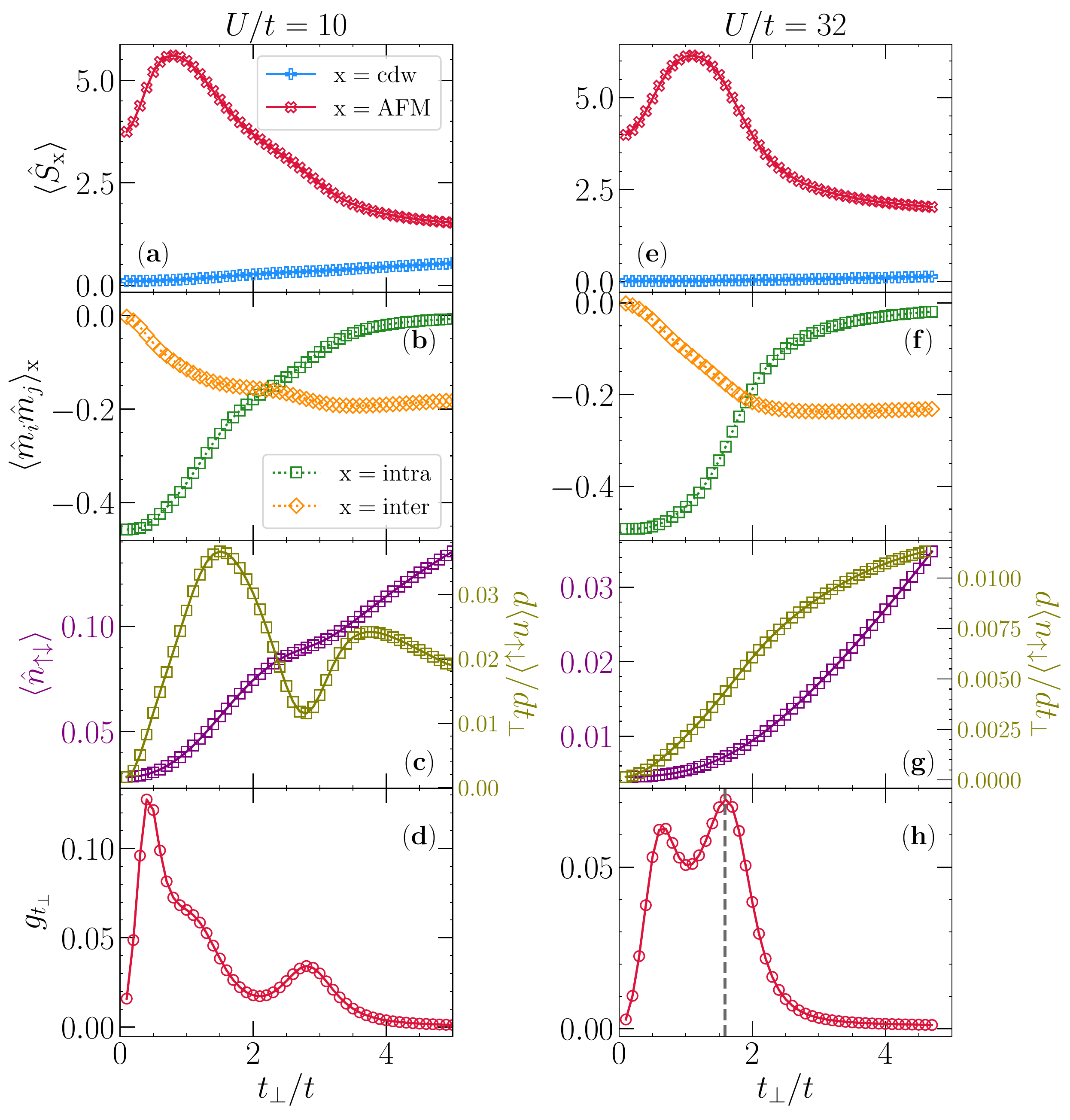}
    \caption{ED results of observables in a $\sqrt{8}\times\sqrt{8}$ bilayer at $U/t=10$ versus interplane hybridization: (a) the expectation value of the staggered charge and spin-structure factor, (b) intra- and interplane spin-spin nearest neighbor correlations, (c) double occupancy, and (d) the fidelity susceptibility. (e)-(h) The same for $U = 32t$; vertical dashed line in (h) depicts the Heisenberg result from Refs.~\cite{Sandvik1994,Wang2006} $J^c_\perp/J=2.52181(3)$ converted to $t^c_\perp/t=1.58802(1)$ via their correspondence in the strongly interacting regime, $J \propto t^2/U$.}
    \label{fig:figS2}
\end{figure}

We start by analyzing the strongly interacting regimes, where finite-size effects are typically less dramatic: Figs. \ref{fig:figS2}(a)-\ref{fig:figS2}(d) for $U/t=10$, and Figs.\ref{fig:figS2}(e)-\ref{fig:figS2}(h) for $U/t=32$. The dependence of the antiferromagnetic spin structure factor (see main text for definition) is shown in Fig.\ref{fig:figS2}(a) and\ref{fig:figS2}(e). As the system evolves from a typical planar antiferromagnet to a bilayer one as $t_\perp$ increases, i.e., the number of neighbors effectively grows, this quantity (computed in the ground-state $|\Psi_0\rangle$) initially increases. However, at larger interplane hybridization, it is then suppressed for $t_\perp \gtrsim t$ as singlets form, similar to the behavior seen in the DQMC results in the main text. Concomitantly, given the reduced spin ordering, the staggered charge structure factor, $S_{\rm cdw} \equiv (1/2L^2)\sum_{i,j} (-1)^{i+j}\langle(\hat n_{i\uparrow} + \hat n_{i\downarrow})(\hat n_{j\uparrow} + \hat n_{j\downarrow})\rangle$ slightly increases with $t_\perp$, albeit with an overall small magnitude, suggesting the absence of a charge density wave formation. The smooth evolution of these two quantities is evidence against the manifestation of a first-order phase transition in the range of parameters studied.

The reduction of global antiferromagnetic order giving way to a unordered state displaying interplane spin singlets can be also seen by directly computing the \textit{nearest-neighbor} intra- and interplane spin correlation functions, $\langle \hat m_i \hat m_j\rangle\equiv \langle(\hat n_{i\uparrow} - \hat n_{i\downarrow})(\hat n_{j\uparrow} - \hat n_{j\downarrow})\rangle$. All such spin correlations are negative, which signals their local antiferromagnetic character. With increasing $t_\perp$, interplane spin correlations surpass intraplane ones in magnitude, preceding the onset of the quantum disordered phase.

The double occupancy, $\langle \hat n_{\uparrow\downarrow}\rangle \equiv  \langle \hat n_\uparrow \hat n_\downarrow\rangle$, in analogy with Figs.~\ref{fig:figS_size}(b) and \ref{fig:figS_size}(d), can give the locations of the insulator-metallic pseudo-(or finite-size influenced) transitions at small interactions, in particular, when its derivative, $d\langle \hat n_{\uparrow\downarrow}\rangle/dt_\perp$, is evaluated. At large $U/t$, Fig.~\ref{fig:figS2}(g), their absolute value is relatively small, and reflects the approach to the Heisenberg regime of localized spins.

While much can be captured by the dependence of few-body correlators as we have done so far, a direct account of how the many-body ground state changes with the increasing interplane hybridization, can be understood in terms of the fidelity susceptibility $g_{\perp} = \frac{1}{L^2}\frac{1-|\langle \Psi_0(t_\perp)|\Psi_0(t_\perp+dt_\perp)\rangle|}{dt_\perp^2}$~\cite{Zanardi06,CamposVenuti07,Zanardi07,You2007}, which can locate a quantum phase transition without making direct assumptions regarding a possible order parameter ~\cite{Yang07,Jia11,Mondaini15}. This quantity displays a peak that is extensive, and independent of the (sufficiently small) parameter `perturbation' for continuous phase transitions. Here, we take $dt_\perp=10^{-3} t$, and we note the presence of a first peak at small $t_\perp$ as related to the crossover from planar to bilayer antiferromagnetism. A second peak is also visible, which in the case of almost fully formed local spins captures the magnetic transition, closely following the results obtained for the Heisenberg model for much larger lattices \cite{Sandvik1994,Wang2006}.

\begin{figure}[t]
    \centering
    \includegraphics[width=1.0\columnwidth]{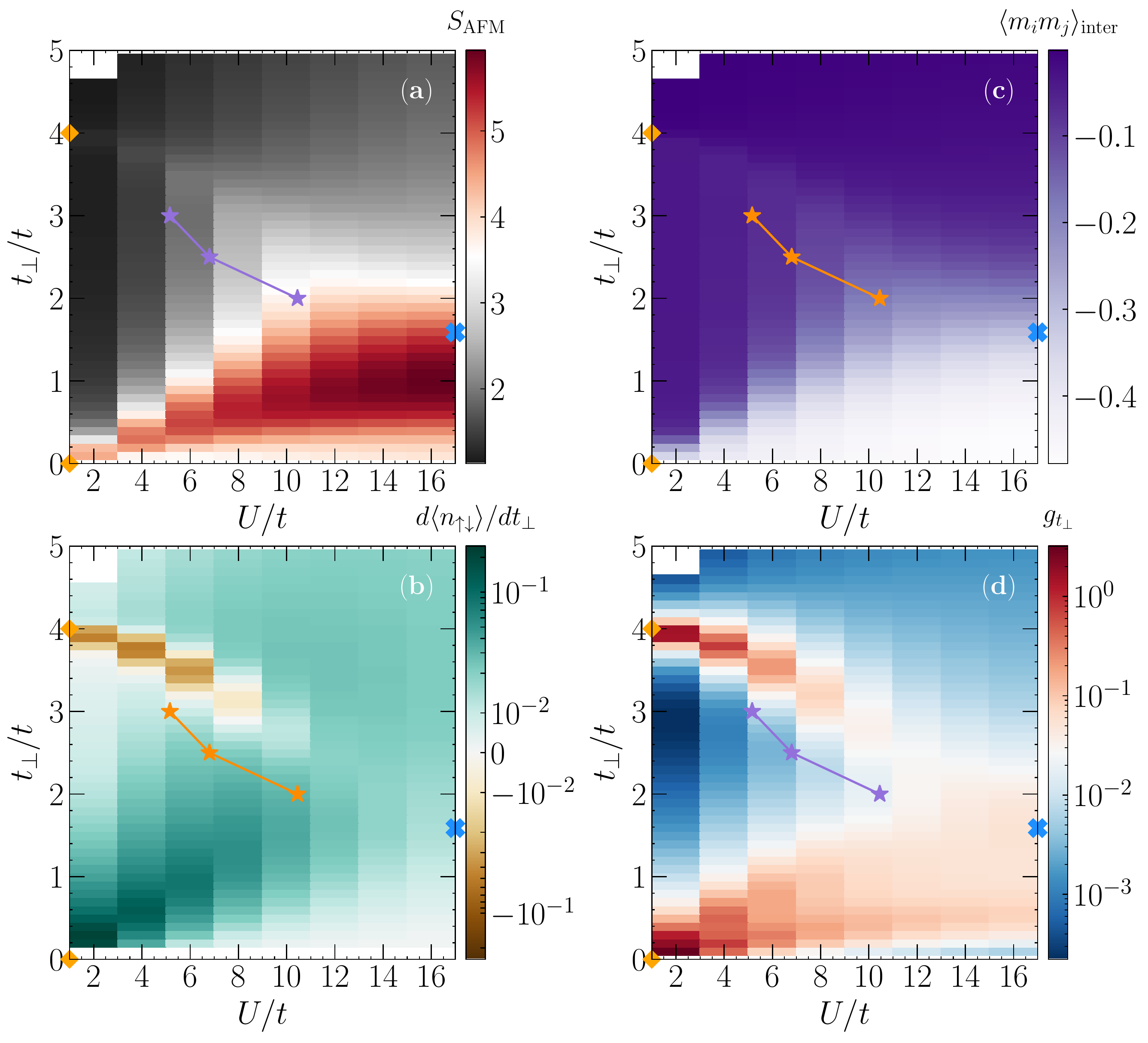}
    \caption{Color maps of several quantities in the space of parameters $t_\perp$ versus $U$ for the $2\times8$-site bilayer lattice. (a) Expectation value of the antiferromagnetic spin-structure factor, (b) derivative of the double occupancy in respect to $t_\perp$, (c) interplane spin-spin nearest-neighbor correlations, and (d) the fidelity susceptibility $g_{t_\perp}$. Star markers depict the magnetic transition at $T=0$ obtained using QMC in much larger lattices in Ref.~\cite{Golor2014}; cross marker on the right (large $U$) panel edge denotes the Heisenberg limit~\cite{Sandvik1994,Wang2006}; and diamond-shaped ones mark the nesting conditions in the noninteracting regime for this cluster size. }
    \label{fig:figS3}
\end{figure}

A more general description of such quantities across the $t_\perp-U$ phase diagram is given in Fig.~\ref{fig:figS3}. At small interactions, the various observables are deeply affected by finite-size effects, manifesting the influence of the metallic region extending at finite values of $U$. In this small cluster, only eight $k$ points are available in the Brillouin zone. In this case, the nesting condition (see main text) between the two bands occurs only at $t_\perp/t=0$ and 4. As a result, the phase diagrams are remarkably similar to the ones obtained in small clusters in DMFT~\cite{Kancharla2007}. In particular, we can highlight (i) the large antiferromagnetic structure factor at small $t_\perp$ denoting the magnetically ordered phase [Fig.~\ref{fig:figS3}(a)]; (ii) the $d\langle \hat n_{\uparrow\downarrow}\rangle/dt_\perp >0$ [$d\langle \hat n_{\uparrow\downarrow}\rangle/dt_\perp<0$] describing the insulator-metal [metal-insulator] transition for this small cluster [Fig.~\ref{fig:figS3}(b)], similar to the results of Fig.~\ref{fig:figS_size};  (iii) the onset of the quantum disordered phase, signified by the close-to-saturation of the interplane nearest-neighbor spin correlations, $\langle \hat m_i \hat m_j\rangle_{\rm inter}$ in Fig.~\ref{fig:figS3}(c); and (iv) the second peak (branch) of the fidelity susceptibility at $t_\perp/t \approx 4$ describes the magnetic transition at small interactions, and converges to the Heisenberg limit at sufficiently large values of $U/t$ [Fig.~\ref{fig:figS3}(d)].

Lastly, we provide a simplified view of the single-particle gap within ED calculations by computing the charge-gap for excitations,
\begin{equation}
    \Delta_c = [E_0(N_e+1) - E_0(N_e)] - [E_0(N_e) - E_0(N_e-1)],
\end{equation}
where $E_0(N_e)$ is the ground-state energy with $N_e$ electrons. Since the lattice is small, this quantity displays relatively large finite-size effects, but in principle should capture similar information as the single-particle gap extracted from QMC simulations in the main text, in particular, in the case where twisted-boundary conditions are applied~\cite{Didier91}.

\begin{figure}[t]
    \centering
    \vspace{0.6cm}
    \includegraphics[width=0.82\columnwidth]{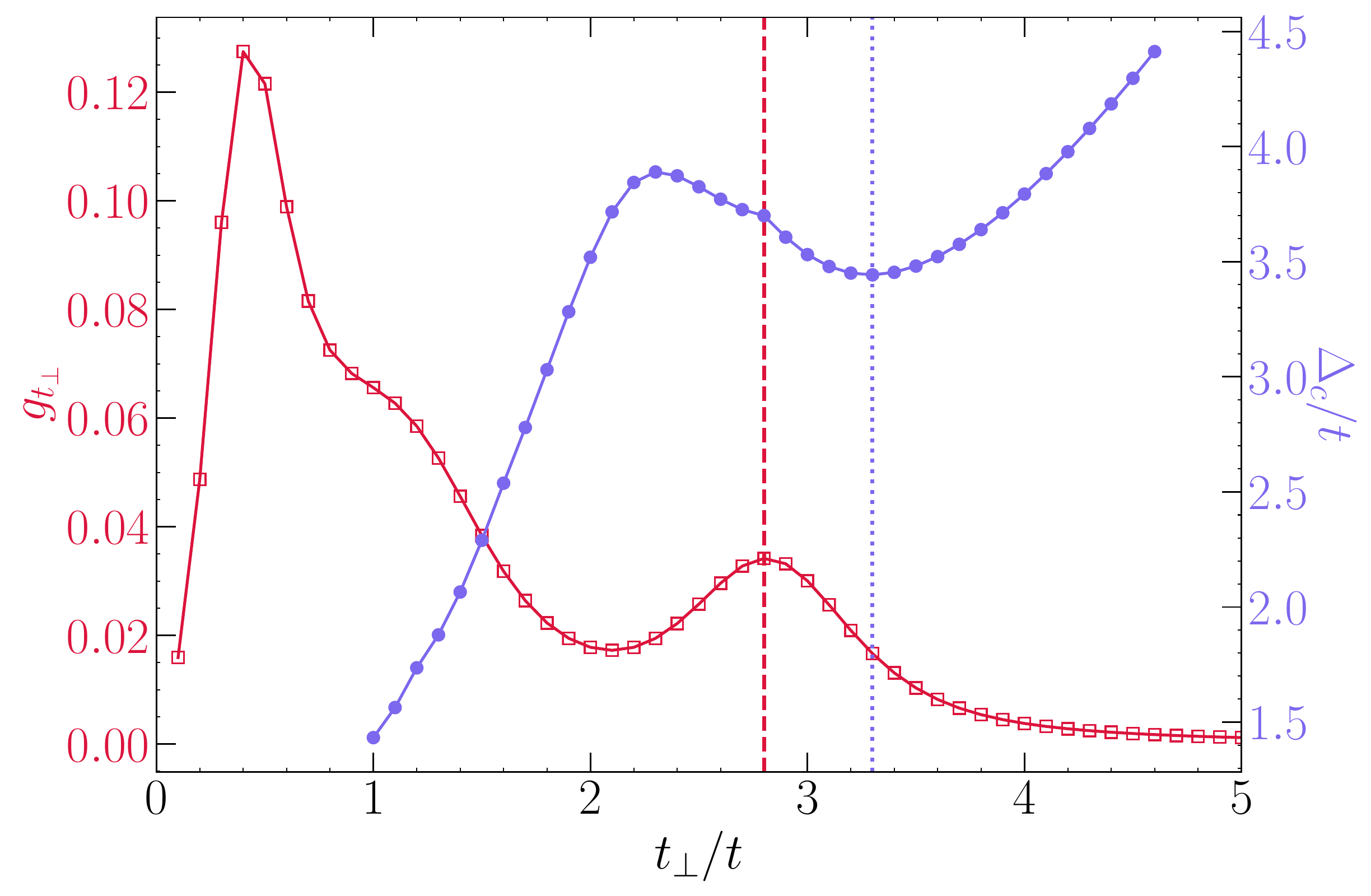}
    \caption{The charge gap $\Delta_c$ (right-axis) versus interplane hybridization for a $\sqrt{8}\times \sqrt{8}$ bilayer, computed via ED with $U/t=10$. For comparison, we also show the fidelity susceptibility (left-axis) at the same interaction strength. The local minimum (maximum) of $\Delta_c$ ($g_{t_\perp}$) is marked via a dotted (dashed) vertical line at $t_\perp/t=3.3$ ($t_\perp/t=2.8$).}
    \label{fig:figS4}
\end{figure}

Here, using standard periodic boundary conditions, we report $\Delta_c$ in Fig.~\ref{fig:figS4}, accompanied by the corresponding fidelity susceptibility. We avoid the regime of small interactions, focusing on $U/t=10$, and contrast it with the results in much larger systems obtained via QMC in Fig.~\textcolor{red}{3}(b) of the main text.

Although at small values of $t_\perp$ the gaps $\Delta_c$ and $\Delta_{\rm sp}$, extracted from ED and QMC respectively, do not display similar behavior (notwithstanding the strikingly different lattice sizes), at large hybridization, both exhibit a minimum at around $t_\perp/t \simeq 3$, which we have associated to a crossover from a paramagnetic Mott insulator towards a band insulator in the main text. To argue that the magnetic transition is not aligned with the above described crossover, we overlay the data originally shown in Fig.~\ref{fig:figS2}(d) for the fidelity susceptibility in Fig.~\ref{fig:figS4}. The peak at large $t_\perp/t$, which we inferred to be related to the magnetic transition, does not coincide with the local minimum of $\Delta_c$, opening room for a potential paramagnetic Mott insulator surviving at $2.8 < t_\perp/t < 3.3$ for this lattice size and for $U/t=10$.

\bibliography{bilayer_bib}

\end{document}